\documentclass[aps,prb,superscriptaddress,twocolumn,nofootinbib]{revtex4-2}
\usepackage[colorlinks,urlcolor=blue,citecolor=blue,linkcolor=blue]{hyperref}
\usepackage{graphicx,amsmath,amsfonts,amssymb,subfigure,dsfont,color}
\usepackage[table]{xcolor}
\usepackage{soul}
\usepackage{bbold}

\usepackage{tcolorbox}
\definecolor{light_blue}{HTML}{f0f5ff}
\definecolor{light_grey}{HTML}{ededed}
\definecolor{mygreen}{HTML}{5ca637}
\definecolor{myblue}{HTML}{12a2db}
\definecolor{mypurple}{HTML}{9e3ac2}
\definecolor{dark_magenta}{HTML}{b30047}

\tcbset{fontupper=\normalsize,
	colback=white, colframe=black, colbacktitle=light_grey,
	coltitle= black, arc=0.25mm, center title}

\newcommand{\thetitle}{Extending the self-discharge time of Dicke quantum batteries using molecular triplets}

\begin{document}
	
\title{\thetitle}

\author{Daniel J. Tibben} 
\author{Enrico Della Gaspera} 
\author{Joel van Embden} 
\author{Philipp Reineck} 
\affiliation{School of Science, RMIT University, Melbourne, VIC, 3000, Australia.}
\author{James Q. Quach} 
\affiliation{CSIRO, Ian Wark Laboratory, Bayview Ave, Clayton, Victoria, 3168, Australia.}
\author{Francesco Campaioli}
\email{francesco.campaioli@unipd.it}
\affiliation{Dipartimento di Fisica e Astronomia ``G. Galilei'' Università degli Studi di Padova, I-35131 Padua, Italy.} 
\affiliation{Padua Quantum Technologies Research Center,  Università degli Studi di Padova, I-35131 Padua, Italy.}
\author{Daniel E. G\'omez}
\email{daniel.gomez@rmit.edu.au}
\affiliation{School of Science, RMIT University, Melbourne, VIC, 3000, Australia.}

\date{\today}

\begin{abstract} 
    Quantum batteries, quantum systems for energy storage, have gained interest due to their potential scalable charging power density. 
    A quantum battery proposal based on the Dicke model has been explored using organic microcavities, which enable a cavity-enhanced energy transfer process called superabsorption. 
    However, energy storage lifetime in these devices is limited by fast radiative emission losses, worsened by superradiance. 
    Here, we  demonstrate a promising approach to extend the energy storage lifetime of  Dicke quantum batteries using  molecular triplet states. 
    We examine a type of multi-layer microcavities where an active absorption layer transfers energy to the molecular triplets of a storage layer, identifying two regimes based on exciton-polariton resonances. 
    We tested one of these mechanisms by fabricating and characterising five devices across a triplet-polariton resonance. 
    We conclude by discussing potential optimisation outlooks for this class of devices.
\end{abstract}

\maketitle

\section{Introduction}
\label{s:introduction}

Batteries, now a crucial element in a wide range of energy-efficient applications, are being seamlessly integrated into smart electronics, textiles, the Internet of Things, and electric vehicles, thereby reshaping our way of life. 
Large-scale battery-based energy storage is proving instrumental in addressing intermittency challenges associated with renewable energy sources such as solar and wind power~\cite{Larcher_NC2015a}. 
However, the current generation of electrochemical batteries, especially Li-ion ones, are largely deficient in their ability to charge quickly and efficiently; they suffer from rapid degradation and require replacement after only a few hundred cycles~\cite{Armand_N2008a}. 
Consequently, the advent of novel energy storage technologies (beyond electrochemistry)
with  fast charging yet maintaining high energy density 
is urgently required~\cite{Ye_NE2024a}.

\begin{figure*}[th!]
	\centering
	\includegraphics[width=0.93\textwidth]{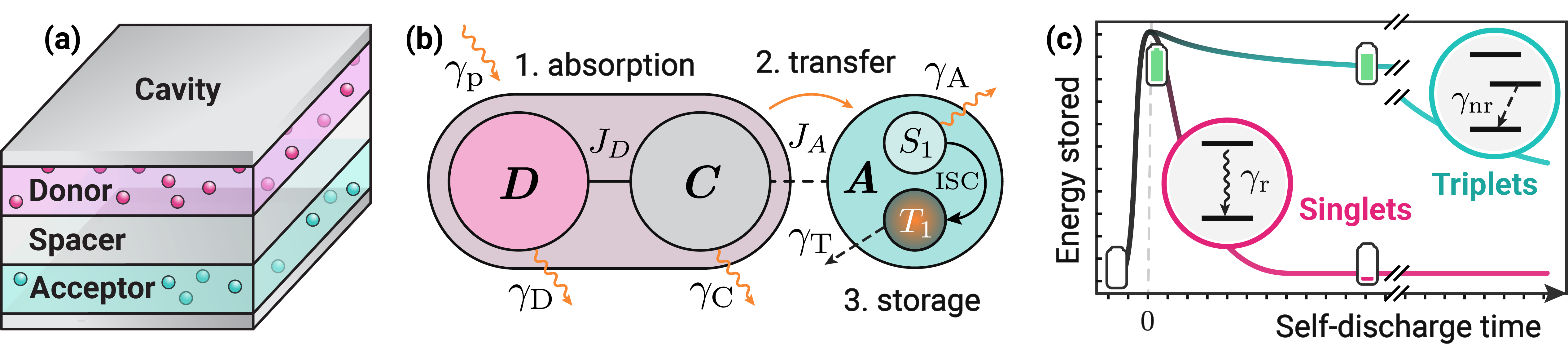}
	\caption{\textbf{Storage layer in a Dicke exciton-polariton quantum battery}---
    (a) We consider a microcavity with a donor and acceptor layer, separated by an inert polymer spacer to avoid direct dipole-dipole couplings between them. The donor is chosen to be a bright, strongly absorbing dye, while the acceptor is chosen to absorb weakly and to have a dark molecular triplet state.
	(b) We use an effective Jaynes-Cummings model describing the optoelectornic properties of the devices, consists of a cavity ($C$) strongly coupled to the donor's singlet ($D$) via $J_D$ and weakly coupled to the acceptor's singlet ($S_1$ at $A$) via $J_A$. The acceptor's triplet state ($T_1$ at $A$) can be populated via intersystem crossing (ISC) from its singlet ($S_1$) or optically via weak dipole-dipole couplings. 
    (c) By transferring energy to the dark molecular triplet states of the acceptor layer we aim to prevent the detrimental effects of superradiance, i.e., the counterpart of superabsorption, which leads to short self-discharge time due to cavity-enhanced radiative emission losses ($\gamma_\textrm{r}$) from the bright donor singlets, usually in the pico to nanosecond timescale. Instead, triplets' non-radiative relaxation ($\gamma_\mathrm{nr}$) can be orders of magnitude slower, leading to micro-second self-discharge times.}
	\label{fig:device_design}
\end{figure*}

In a quantum battery (QB), a theoretical construct for energy storage, energy is stored in the excited states of a quantum system, such as an atom or a molecule, and can be extracted by inducing transitions between these states. QBs aim to harness properties like superposition and quantum entanglement to achieve a \textit{quantum advantage} for energy storage: Much alike Grover's search algorithm for quantum computing~\cite{leuenberger2001quantum}, quantum batteries can offer a $\sqrt{N}$-fold charging speed-up that grows with the number $N$ of sub-cells involved~\cite{Campaioli_2023a}. 
A promising architecture of a QB is the \textit{Dicke quantum battery}, proposed by~\citet{Ferraro_PRL2018}, which is based on the Dicke model of light-matter interaction between an optical cavity and a set of (non-interacting) emitters~\cite{RevModPhys.85.1083,cong2016dicke,fusco2016work}. 
~\citet{Ferraro_PRL2018} theoretically demonstrated  that a Dicke quantum battery's charging power density could scale with the number of optical emitters  $N$ as  $\sqrt{N}$, by leveraging a phenomenon known as \textit{superabsorption}~\cite{higgins2014superabsorption,Erdman2022,Crescente2020,Dou2022,Gemme2023a}.
A significant breakthrough was achieved with the recent experimental demonstration of superabsorption on an organic optical microcavity~\cite{Quach_SA2022a}. 
The results suggest that the energy transfer rate from the optical cavity mode to the photoactive molecules in the microcavity (analogous to a battery charging speed), increases superextensively, thereby supporting the superabsorption hypothesis: 
QBs charge faster as their storage capacity increases.

Superabsorption, however, is a double-edged sword. 
While it enables ultra-fast charging, i.e., the transfer of energy from the cavity to the emitters, it also introduces its well-known counterpart, \textit{superradiance}, which results in a cavity-enhanced relaxation pathway, leading to fast battery discharge~\cite{Chen_PRL2013a,Crisp_NL2013a,Svidzinsky_PRX2013a,Aberra-Guebrou_PRL2012a,Garraway_PTOTRSAMPAES2011a,Prasad_PRA2010a,Temnov_PRL2005a,Prasad_PRA2000a,Gross_PR1982a,Dicke_PR1954a}.
Although superradiance can be partially offset by local decoherence~\cite{Quach_SA2022a}, the natural radiative emission rate of individual emitters (typically nanoseconds) sets an upper--limit on the storage time of that version of a Dicke QB. 
Indeed  \citet{Quach_SA2022a} observed extremely brief self-discharge time (nanoseconds). 
Therefore, a  challenge in this research area is to discover strategies to prolong the storage lifetime of such devices, a critical prerequisite for their prospective application in energy storage.

Here we propose a Dicke QB architecture where superextensive absorption is spatially decoupled to energy storage within the same device. 
To achieve this, the devices consist of donor and acceptor layers, the former responsible for superabsorption while the latter has a built-in mechanism that permits prolonged energy storage.
As an embodiment of the concept, the acceptor layer we study in this letter stores energy using molecular triplet states (or triplet \textit{excitons}). 
Molecular triplet states ($T_1$, Fig.~\ref{fig:device_design}) are typically orders of magnitude darker than the bright singlet states ($S_1$, Fig.~\ref{fig:device_design}), while being only slightly less energetic, thus offering a promising approach to counter the \textit{detailed-balance} imposed by the superabsorption-superradiance ``duality''. 
While dark electronic states have been considered to reduce losses in energy harvesting~\cite{tayebjee2017quintet,collins2023quintet}, exciton transport~\cite{Davidson2020,Davidson2022} and energy storage~\cite{Liu2019,Quach2020}, so far, dark states have been produced mostly  through superpositions of locally bright states, e.g., the ground states of H-aggregates~\cite{Hestand2018}. 
These, however, tend to be sensitive to local disorder and decoherence that perturb the symmetry required to cancel out optical dipole moments~\cite{Liu2019}. 
Here, instead, we focus on molecular triplet states, which are naturally prohibited to relax via radiative emission due to spin-forbidden transitions~\cite{Yost2012}, leading to exciton lifetimes that span from the microseconds to the minutes~\cite{liang2020ultralong}, and therefore longer energy storage times in QBs (Fig 1(c)).

To this end, our proof-of-concept device is based on a  multi-layer optical microcavity, schematically shown in Fig.~\ref{fig:device_design}, wherein the interaction between the electromagnetic mode of the cavity and an ``donor"  layer ensures superabsortion~\cite{Quach_SA2022a} (i.e. fast battery charging).
The absorbed energy is transferred  into an acceptor \textit{storage layer} comprised of molecular species with efficient interystem crossing and  with long--lived triplet states, therefore preventing superradiant battery discharge. 
Using an effective Jaynes-Cummings model, we show  two possible pathways to store energy in our architecture, both mediated by resonance conditions, and characterised by widely different optoelectronic and energy storage properties.
We experimentally realise these devices and show that energy is stored for tens of microseconds, which would lead to a $\sim 10^3$--fold increase in the storage time of a Dicke quantum battery.

\section{Storing energy in molecular triplet states}
\label{s:storing_energy_in_triplets}

\subsection{Device design and modelling}
\label{ss:device_design}

\begin{figure*}[t!]
	\centering
	\includegraphics[width=0.93\textwidth]{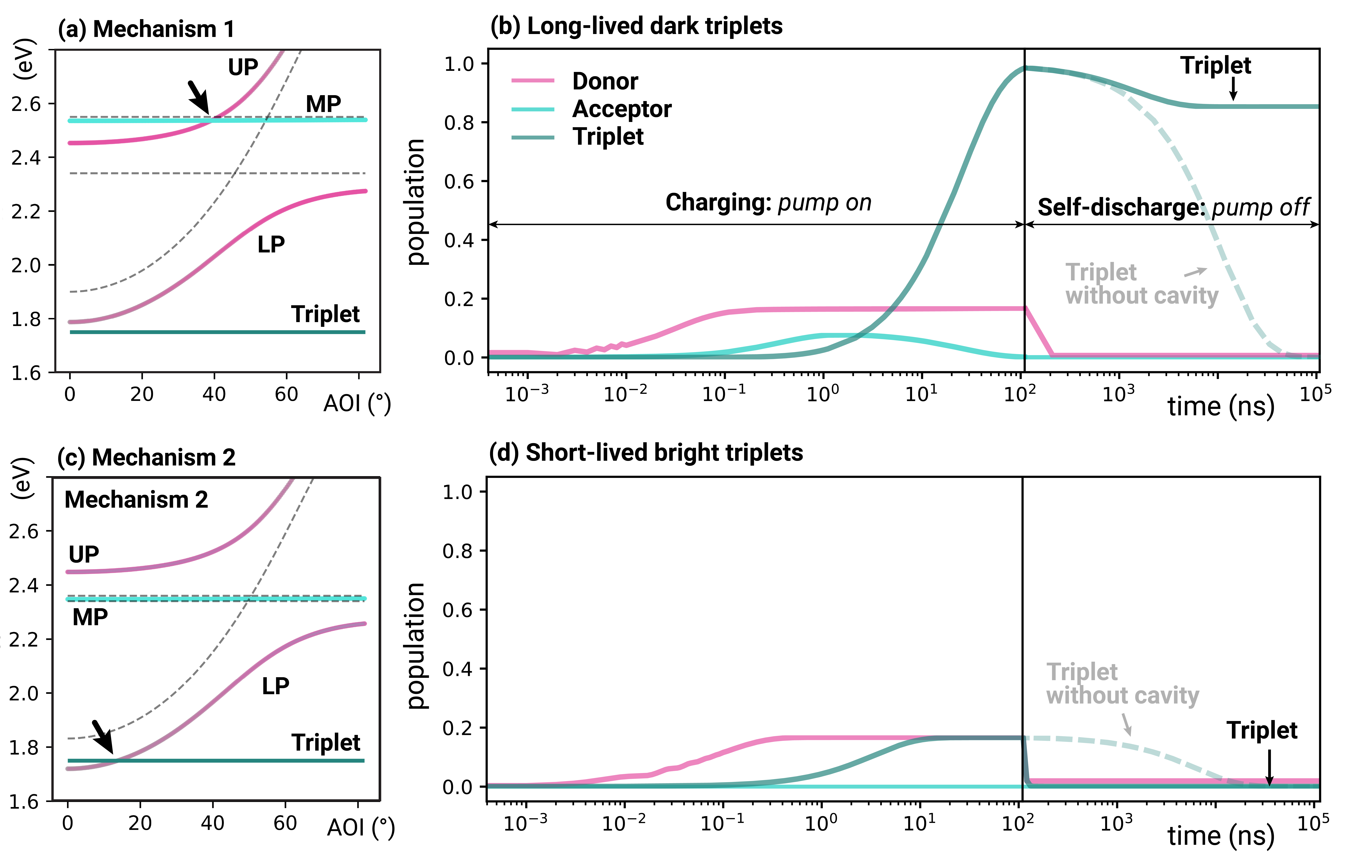}
	\caption{\textbf{Energy storage mechanisms}---Polaritons of the effective Jaynes-Cumming Hamiltonian of Eq.~\eqref{eq:Jaynes-Cummings_RWA} and acceptor's triplet. The colours schematically represent the relative population of each diabatic state, with the middle polariton (MP) being predominately populated by the acceptor's singlet. (\textit{a}) Dark triplets can form via ISC at the resonance between the acceptor's singlet (here, almost exactly given by the ``middle'' polariton) and a bright highly-absorbing polariton (here the ``upper'' polariton, which is a cavity-donor hybrid). In this case the molecular triplets are off-resonance and do not couple with any other state. The triplets formed this way are dark and long lived. (\textit{b}) Bright triplets can form via direct dipole-dipole coupling with a resonant bright polariton (here the ``lower'' polariton LP). The resulting triplets are hybrids (coherently mixed with the polariton state), therefore their lifetime is shortened by radiative recombination. The system parameters used to generate these figures are reported in Tab.~\ref{tab:device_parameters_SM} in Sec.~\ref{sm:oqs_model_dynamics} of the SM.
    (\textit{c})\textbf{Mechanism 1 (\textit{Top}).}---The triplets formed in this case cannot fully decay via IC. A significant portion of the triplet remains populated indefinitely in the absence of additional relaxation pathways, leading to excellent energy storage performance. (\textit{d}) \textbf{Mechanism 2 (\textit{Bottom}).}---The bright hybrid triplets formed this way decay as fast as the cavity-donor polaritons, leading to poor energy-storage performance.}
	\label{fig:mechanisms}
\end{figure*}

Our objective in this section is to establish conditions that will enhance the efficiency of energy transfer from an external driving field, such as a continuous-wave (CW) laser, to the energy storage layer of Fig.~\ref{fig:device_design}.
To this end,  we use an effective Jaynes-Cummings Hamiltonian associated with the underlying many-body Dicke model\footnote{In the Dicke model each donor and acceptor molecule in the ensembles is treated as an individual system.}~\cite{Farina2019,carlonzambon2020,Tibben_CR2023} in the rotating-wave approximation\footnote{When the exciton-photon couplings are small enough compared to the other energy scales, i.e., $J_D,J_A \ll \omega_C(\theta),\omega_D,\omega_A$, we can apply the rotating wave approximation to eliminate the counter rotating terms arising from the exciton-cavity couplings.},
\begin{equation}
	\label{eq:Jaynes-Cummings_RWA}
	H = \omega_C(\theta)a^\dagger a + \sum_{\nu = D,A} \omega_\nu \sigma_+^{(\nu)}\sigma_-^{(\nu)} + J_\nu \sigma_+^{(\nu)}a + \sigma_-^{(\nu)}a^\dagger,
\end{equation}
where the cavity mode has an energy $\omega_C(\theta)$ that depends on the angle of incidence of light $\theta$ (note that for shorthand notation we have set $\hbar$ = 1). 
$J_D = \sum_{j=1}^{N_D}$ ($J_A = \sum_{j=1}^{N_A}$) is the cumulative coupling between cavity and donor (acceptor), $\omega_D$ ($\omega_A$) is the average energy\footnote{This represent the energy gap between ground $S_0$ and excited $S_1$ singlet, also known as the \textit{HOMO-LUMO} gap~\cite{zade2006oligomers}.} of the molecules in the donor (acceptor) layer, $a$, $a^\dagger$ are the annihilation and creation operators of the cavity and $\sigma_\pm^{(\nu)} = \sigma^{(\nu)}_x\pm\sigma^{(\nu)}_-$ are the singlet-exciton creation and annihilation operator at chromophore $\nu = D,A$. 
The acceptor $A$ is a three-level system with a  triplet state $T_1$. 
The latter may have a small residual optical dipole moment, which leads to a cavity-triplet coupling with strength $J_T$. 
All incoherent transitions, such as radiative and non-radiative emission losses are modelled using a Markovian quantum master equation. 
This includes a laser pumping rate $\gamma_\mathrm{p}$, a cavity loss rate $\gamma_\mathrm{C}$, singlet exciton radiative emission rates $\gamma_\mathrm{D}$, $\gamma_\mathrm{A}$ for donor and acceptor, respectively, a triplet non-radiative decay $\gamma_\mathrm{T}$, and an intersystem crossing (ISC) rate $\gamma_\mathrm{ISC}$ for the $S_1 \to T_1$ transition at the donor (See Sec.~\ref{sm:oqs_model_dynamics} of the Supplemental Material (SM) for the modelling details).

The strong interaction between the optical cavity mode and the excited (singlet) states of the D and A materials results in the formation of polariton states~\cite{Ebbesen_AOCR2016a}.
In our model, 
we consider cases where the cavity-donor interaction strength greatly surpasses the equivalent cavity-acceptor coupling.
In turn, this translates into 
 \textit{upper} (UP) and \textit{lower} (LP) polaritons with strong cavity-donor hybridisation, and  \textit{middle polaritons} (MP) with nearly 100\% acceptor character. 
 Our model allows us to monitor the impact of parameters that can be experimentally manipulated on the population of triplet states.
This, in turn, enables us to optimise the energy storage within QBs.
We now discuss the two main mechanisms that allow energy transfer to the storage layer of a theoretical QB (Fig.~\ref{fig:mechanisms}).
\begin{figure*}[t!]
	\centering
\includegraphics[width=0.93\linewidth]{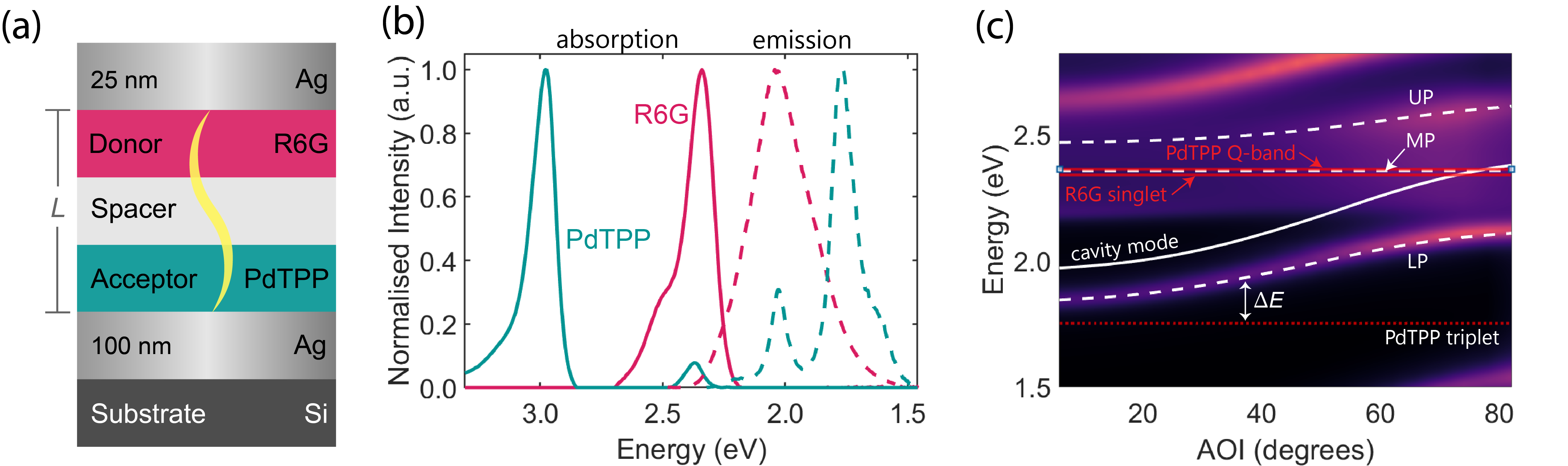}
	\caption{\textbf{Device architecture and properties}---(a) The donor (R6G) is chosen to be a bright, strongly absorbing dye, while the acceptor (PdTPP) is chosen to absorb weakly and to have a dark molecular triplet state.  These are placed in a resonant optical cavity of height \textit{L}. (b) Solid line: Optical absorption spectra of R6G (red) and PdTPP (blue-green). R6G shows a clear vibronic progression around its main singlet excited state at $\sim$2.34 eV. PdTPP shows a strong Soret band at $\sim$2.98 eV with a much weaker Q-band at $\sim$2.36 eV. The weaker Q-band is hybridised with the cavity and R6G to facilitate hybrid mode energy transfer between all 3 components. Dashed line: Emission spectra of R6G (red) and PdTPP (blue-green). The acceptor emission exhibits a distinct emission band centred at $\sim$1.75 eV which originates from its triplet state. (c) Representative absorption dispersion the devices, clearly demonstrating strong coupling. }
	\label{fig:device_structure}
\end{figure*}

\subsection{Mechanism 1: ISC-mediated transfer at polariton-acceptor (singlet) resonance}
\label{ss:mechanism_1}
In this mechanism, the device is superextensively charged due to the collective and synchronised interaction between the cavity mode and the donor singlet states. 
Energy is transferred to the acceptor singlet states whenever there is a resonance condition occurs between the acceptor singlets and a cavity-donor polariton (Fig.~\ref{fig:mechanisms} (a)).
Fast ISC populates the triplet states of the storage layer (Fig~\ref{fig:mechanisms} (c)), which are assumed to have a negligible optical dipole moment and are consequently completely dark. 

Mechanism 1 is efficient if: 
(\textit{i}) the ISC rate $\gamma_\mathrm{ISC}$ is comparable to or faster than the energy loss rates of the device (in turn dictated  by an interplay of singlet donor $\gamma_\mathrm{D}$ acceptor $\gamma_\mathrm{A}$ and cavity photon $\gamma_\mathrm{C}$ decay rates) and 
(\textit{ii}) a resonance condition occurs between the acceptor singlets and a cavity-donor polariton, such as the UP state shown in Fig.~\ref{fig:mechanisms} (a), or equivalently whenever:
\begin{align}
    \label{eq:mechanism_1_1}
    &\gamma_\mathrm{ISC} > \gamma_\mathrm{C}, \gamma_\mathrm{D}, \gamma_\mathrm{A}, \\
    \label{eq:mechanism_1_2}
    &\omega_A \approx E_k,
\end{align}
where $E_k$ is the energy of a (bright) polariton state of the device (see Fig.~\ref{fig:mechanisms} (a), $k$ = UP, LP or MP).
These conditions can be met by carefull choice of materials, where we note that ISC can be ultrafast in some compounds~\cite{zobel2018mechanism,zugazagoitia2008ultrafast,okada1981ultrafast}.
In this mechanism, the triplet states do not mix with singlets or cavity photons, and consequently, their lifetime remains unchanged and set the time scale for energy storage (Fig.~\ref{fig:mechanisms}).
\subsection{Mechanism 2: Optically-driven transfer at polariton-triplet resonance}
\label{ss:mechanism_2}
Following superextensive absorption by the donor layer in the device, energy can be directly transferred to the triplet states of the storage layer whenever one of the polariton states is resonant with the triplet state, and these states have a non-negligible \textit{singlet character}.
In layers with moderate to high dye concentration, random molecular aggregation can lead to triplets with singlet character~\cite{sternlicht1963triplet,atkins1975magnetic,forecast2023magnetic}, namely, with an effective overall spin 0 and a non-vanishing dipole moment.
This leads to triplets that can interact with the cavity via a small dipolar coupling $J_T \ll J_D, J_A$, and can therefore be populated directly via a polariton-triplet resonance, as indicated in Fig.~\ref{fig:mechanisms} (b) and (d).
The resulting triplet states $\widetilde{T}$ can form via coherent mixing between the bare acceptor triplets and the cavity-donor polariton, therefore their lifetime $\tau_{\widetilde{T}}$ is shorter than then bare triplet lifetime $\tau_{T} = \gamma_\mathrm{T}^{-1}$, and is approximately given by
\begin{equation}
	\label{eq:triplet_lifetime}
	\tau_{\widetilde{T}} \approx p_T \tau_T + p_{P} \tau_P,
\end{equation}
where $p_T$ and $p_P$ is the population of the bare triplet and polariton states, respectively, and $\tau_P$ is the lifetime of the polariton state in resonance with the triplet state.

\section{Results}
\label{s:results}

To explore the potential of the triplet storage layer, we created the devices illustrated in Fig.\ref{fig:device_structure} (a), via  thin film deposition, as detailed in Sec. \ref{sm:fabrication_details} of the SM. 
We chose \textit{Rhodamine 6G} (R6G) for the donor layer due to its robust optical absorption in the visible range,
and small ISC quantum yield ($< 10^{-2}$)~\cite{Dempster_JOP1973a}
while the acceptor layer is made of \textit{Palladium tetraphenylporphyrin} (PdTPP), due to its efficient ISC to its molecular triplet state (ISC quantum yield: 0.96 $\pm$ 0.04~\cite{Rogers_TJOPCA2003a}) and weak visible absorption (at its Q-band). 
The optical and electronic properties of this donor/acceptor pair are presented in Fig.\ref{fig:device_structure} (b) and (c). 
We designed the total distance L between the Ag mirrors to generate a second-order Fabry-P\'erot cavity mode with anti-nodes at the locations of the donor and acceptor layers (wavy line in Fig.~\ref{fig:device_structure} (a)).
The energy difference $\Delta E$ between the LP and the triplet state defines the detuning $\Delta E$ of the microcavity, $\Delta E = E_\mathrm{LP} - E_\mathrm{T_1}$ (Fig. \ref{fig:device_structure} (c)).
(associated Jaynes-Cummings model parameters are summarised in Tab.~\ref{tab:device_parameters}).
In our experiments, we monitor the energy storage of each device by measuring their steady-state and time-resolved fluorescence and phosphorescence, where the latter is a directly correlated with the population of triplet states in the storage layer.

\subsection{Steady-state emission}
\label{ss:steady-state_ emission_intensity}

\begin{figure*}[ht!]
   \centering
    \includegraphics[width=\textwidth]{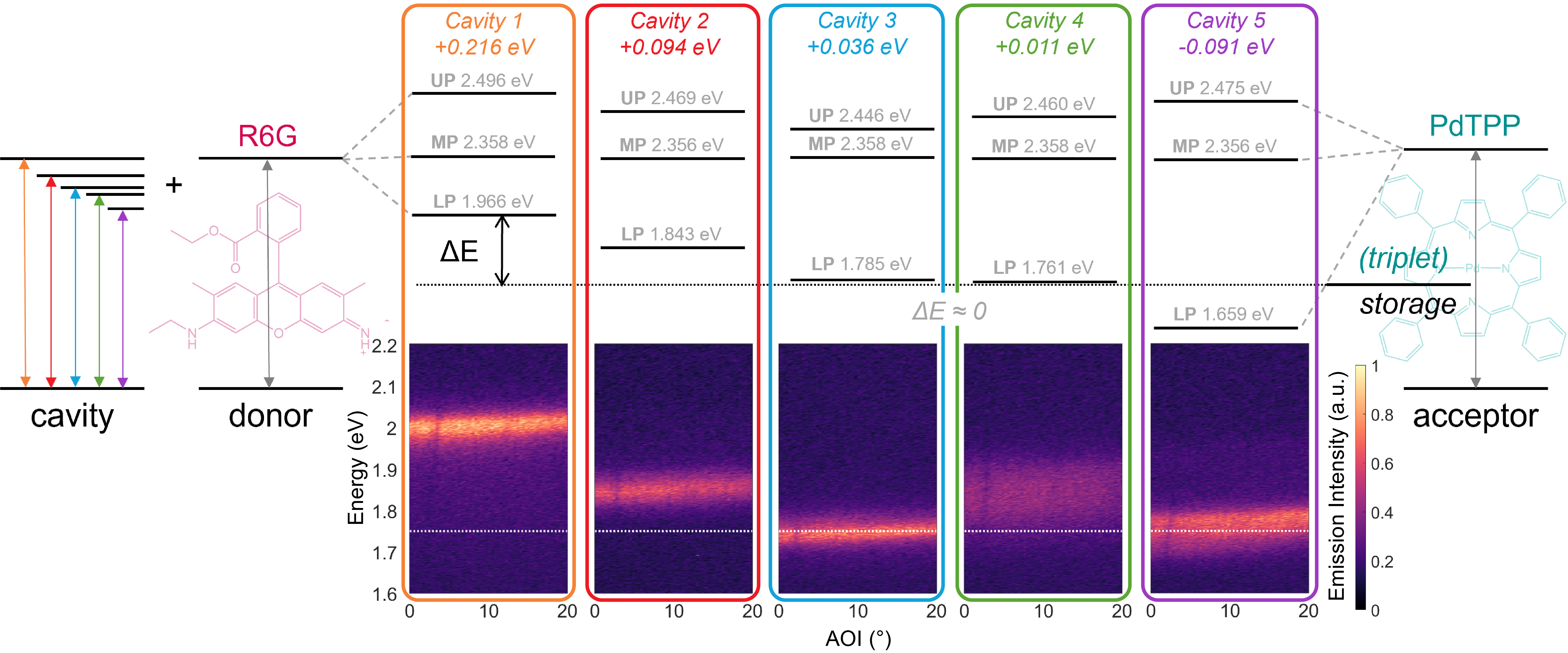}
   \caption{\textbf{Fluorescence emission properties across the LP-Triplet resonance}--- Energy diagram outlining cavity detuning across the devices, with respective polariton energies from the effective Jaynes-Cummings model of Eq. \ref{eq:Jaynes-Cummings_RWA}. Labelled upper (UP), middle (MP), and lower (LP) polariton. Fluorescence emission dispersions are normalised to Cavity 1. As $\Delta E \rightarrow 0$, emission intensity is reduced and its features broaden about the triplet energy (white dashed line), exemplified in Cavity 4. We attribute this to an increase in population of the triplet state due to the optically-driven mechanism of Sec. \ref{ss:mechanism_2}, resulting in reduced brightness in the device.}
   \label{fig:results}
\end{figure*}

As shown through angle-dependent reflectometry in Fig.~\ref{fig:absorption-emission} of the SM, our devices operate in the strong light-matter coupling regime, as evidenced by anti-crossing of the absorption bands around the intersection of exciton  and cavity energies.
These measurements were accurately described with a three-level coupled oscillator model, as detailed in Sec~\ref{sm:measurments}, 
resulting in the set of parameters shown in  Tab.~\ref{tab:device_parameters}.

\begin{table}
		\caption{Device parameters}
	\begin{tcolorbox}[tabulars*={\renewcommand\arraystretch{1.2}}%
		{c|c|c|c|c},adjusted title=flush left,halign title=flush left,
		boxrule=0.5pt,title = {\textbf{Jaynes-Cummings model device parameters}}]
		\hspace{2.5cm}  & $\omega_C$ (eV) & $J_D$ (eV) & $J_A$ (eV) & $\Delta E$ (eV)  \\
		\hline\hline
		{\color{orange}\textbf{Cavity 1}} & 2.12 & 0.23 & 0.07 & 0.216 \\
		\hline
		{\color{red}\textbf{Cavity 2}} & 1.97 & 0.23 & 0.10 & 0.094 \\ 
		\hline
		{\color{myblue}\textbf{Cavity 3}} & 1.89 & 0.23 & 0.07 & 0.036\\
		\hline
		{\color{mygreen}\textbf{Cavity 4}} & 1.88 & 0.25 & 0.08 & 0.011 \\ 
		\hline
		{\color{mypurple}\textbf{Cavity 5}} & 1.79 & 0.27 & 0.13 & $-$0.091 \\ 
		\hline
	\end{tcolorbox}
	\label{tab:device_parameters}
\end{table}

In Fig.~\ref{fig:results} we show how the emission of light  from the devices (in turn related to their self-discharge) is affected by $\Delta E$. 
In these measurements, the devices were illuminated using 514 nm CW laser irradiation at $\sim1.6 \mu$W, and the emission was measured as a function of the angle of emission of light, relative to the surface normal of each device.

For Cavity 1 and Cavity 2, light is emitted at the spectral location of the LP band with a dependence on angle that matches the one observed for absorption.
This, together with the absence of clear triplet emission, indicates minimal energy transfer from the LP to the triplet at these $\Delta E$ values.
In the isoenergetic ($\Delta E  \approx 0 $) cases of Cavity 3 and Cavity 4, emission occurs at the  triplet energy.
In Cavity 3, this emission is narrow and well-defined about the LP and triplet energy, akin to that observed in Cavity 1 and Cavity 2. 
However, Cavity 4 shows a broadened emission about the LP and triplet energy, and a significant dip in emission intensity. 
We propose this to be a signature of efficient energy transfer to triplet states, as this excited-state energy is stored in the dark triplet (\textit{storage}) states.
Cavity 5, the most negatively-detuned microcavity, shows emission starting at the triplet state and broadening to the LP energy.
These observations are summarised in Fig.~\ref{fig:data_fit}(a) and (b) where we plot the emission intensity and linewdith (normalised with respect to cavity 1) as a function of $\Delta E$.
Using the open quantum system model based on the Jaynes-Cummings Hamiltonian discussed in Sec.~\ref{sm:oqs_model_dynamics} of the SM, it is possible to fit the measured relative change in the fluorescent emission intensity if the transfer of energy to the storage layer occurs via mechanism 2. 
This fit results in  an estimate for the weak triplet-cavity coupling $J_T$  of $\approx 5$~meV, as shown in Fig.~\ref{fig:data_fit}.

\subsection{Time-resolved emission}
\label{ss:time-resolved_ emission_intensity}
It is important here to make the clear distinction between emission intensity (which is dominated by \textit{fluorescence}), and \textit{phosphorescence} intensity. 
We find that fluorescence-dominated emission decreases as $\Delta E \rightarrow 0$, since LP-triplet hybridisation reduces emission brightness at short timescales (i.e. $<$ 15 ns).
Instead, when looking at \textit{phosphorescence} emission at longer timescales (i.e. $>$ 50 ns), we expect to see an increase in the phosphorescence emission and  decay rate as we approach $\Delta E \rightarrow 0$, since the triplets become increasingly mixed with the LP.
This phosphorescence is much lower in intensity compared to fluorescence-dominated emission, persisting over long timescales. In this time scale, we can isolate the time resolved decay of phosphorescence, which is a proxy for the decay of triplets, and thus, a measure of their population. 

With this in mind, we now consider time-resolved emission for the microcavity devices.
``Short" timescale measurements were acquired at a temporal resolution of 0.05 ns, and ``long" timescale measurements were acquired at a temporal resolution of 12.8 ns.
Both short- and long-timescale emission were stimulated with pulsed 520 nm laser irradiation at $\sim$50 nW.

Polariton emission is expected to occur on a picosecond timescale \cite{Wang_JPCL2014}, beyond the resolution of our instrumentation.
R6G  emission lifetime outside the cavity was measured to be $1.48 \pm 0.01$ ns, compared with a lifetime inside the cavity of $1.10 \pm 0.02$ ns.
Due to inefficient ISC in R6G, the decrease in fluorescence lifetime  is assigned to the short lifetime of polaritons: due to strong cavity-R6G coupling, the molecule can radiate light faster.

PdTPP fluorescent emission lifetime outside the cavity was measured to be $1.7 \pm 0.1$ ns, compared with a lifetime inside the cavity of $1.6 \pm 0.1$ ns.
Outside the cavity, the phosphorescent lifetime of PdTPP was measured to be $40.0 \pm 0.4$ $\mu$s, compared with a lifetime inside the cavity of $57 \pm 2$ $\mu$s.

\begin{figure}[t!]
    \centering
    \includegraphics[width=0.48\textwidth]{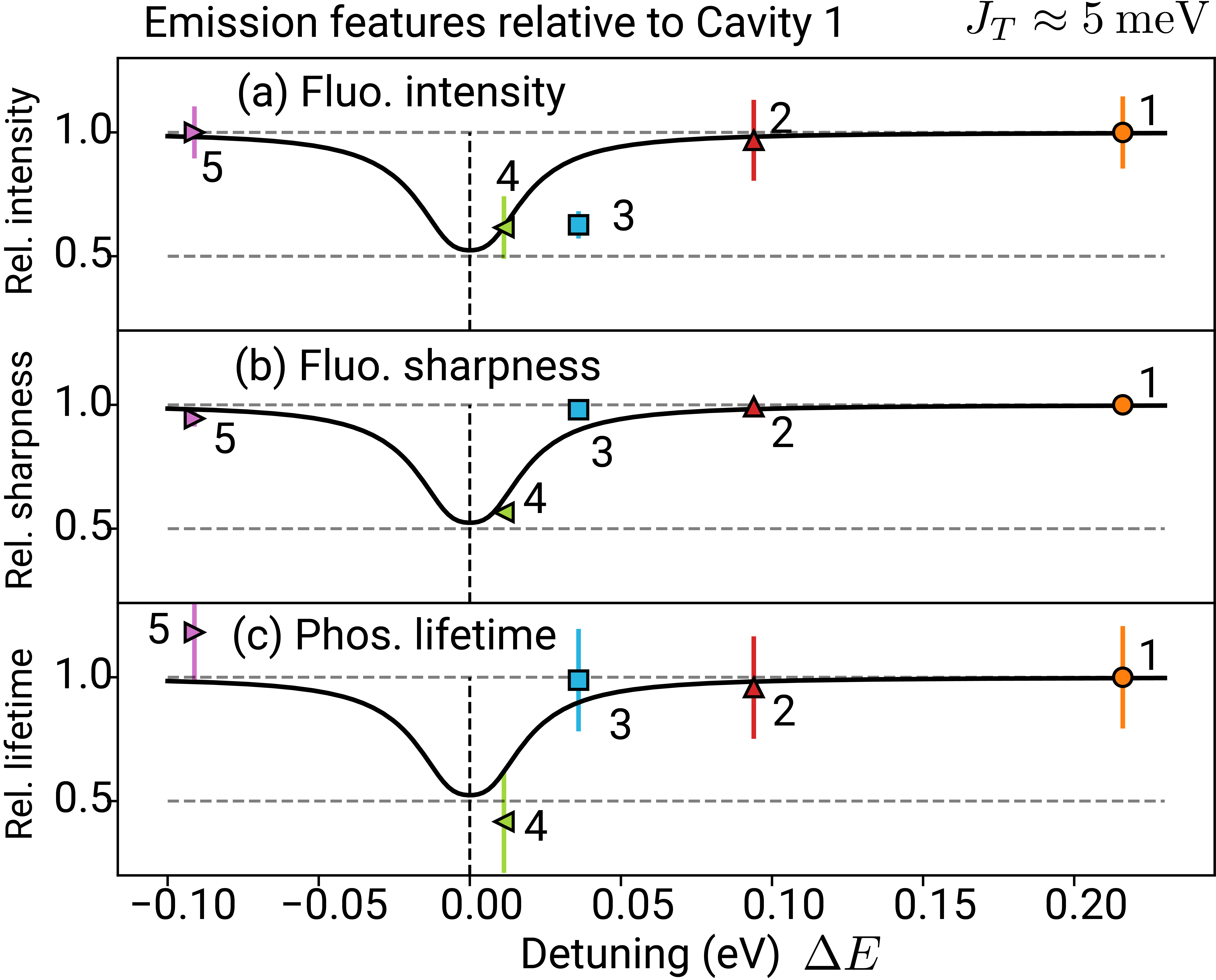}
    \caption{\textbf{Triplet-LP resonance signature via fluorescence and phosphorescence emission}---Relative change in the fluorescence and phosphorescence emission features, with reference to Cavity 1. Fluorescence intensity\footnote{The fluorescent intensity of each device has been normalised to its fluorescent intensity without the cavity, i.e., without the top mirror.} (a) and sharpness (b), here defined as the inverse of the emission broadness (1/FWHM of the emission peaks) are expected to decrease around $\Delta E = 0$ as the LP (fluorescent emitter) is depopulated in favour of the triplets. Conversely, the decay rate of phosphorescence, proportional to the inverse of the triplet lifetime (c), is expected to increase around $\Delta E = 0$, as the molecular triplets (phosphorescent emitters) mix with the LP. We fit these relative features using the model discussed in Sec.~\ref{sm:oqs_model_dynamics} of the SM to obtain a qualitative estimate of the triplet-cavity coupling $J_T \approx 5~\:\textrm{meV}$.
    }
    \label{fig:data_fit}
\end{figure}

Fig.~\ref{fig:data_fit} compares the the time-resolved phosphorescence intensity for each device, normalised to Cavity 1. 
As $\Delta E\to 0$, the phosphorescence decay rate $\gamma_\textrm{ph}$ increases, leading to a shorter triplet lifetime as the latter hybridises with the LP. 
These results are shown in Fig.~\ref{fig:data_fit}, where the phosphorescence ``lifetime'' ($\gamma_\textrm{ph}^{-1}$) dips for Cavity 4, around $\Delta E = 0$, by about 50\% with respect to the other devices. 
Thus, we deduce that, at resonance, the molecular triplet state is populated at around 50\%, resulting in a storage of energy with a self-discharge time of 57 $\pm$ 2 $\mu$s, a clear demonstration of extended energy storage.
This self-discharge time is  at least three orders of magnitude longer than associated with the devices studied by~\citet{Quach_SA2022a}, which are limited by Lumogen-F orange (LFO) radiative emission rate of about $1$~ns.

From these results we can estimate energy density and self-discharge time of the device. 
The energy density $\rho_E = p_\mathrm{T_1} \cdot E_\mathrm{T_1} \cdot \rho_\mathrm{T_1}$, is proportional to the fraction of excited molecular triplets $p_\mathrm{T_1}$, their energy $E_\mathrm{T_1}$ and their density in the device $\rho_\mathrm{T_1} \approx 3.01\times10^{18} cm^{-3}$.

Assuming that at resonance $p_\mathrm{T_1} \approx 0.5$ leads to an energy density of around $\rho_E \approx 2.63\times10^{18} eV cm^{-3}$. 
The total battery capacity $C$, achieved when the triplet states are fully populated, is approximately 
$C = E_\mathrm{T_1} \cdot n_\mathrm{T_1}\approx  1.23\times10^{14} eV$ $\approx 5.47\mu$Wh, where $n_\mathrm{T_1}$ is the total number of PdTPP molecules in the device. 

Further energy storage improvements are possible with our device architecture.
In PdTPP, although ISC is almost perfectly efficient, it occurs on the nanosecond timescale~\cite{Rogers_TJOPCA2003a}, which is comparable to the radiative losses of the cavity and the radiative decay of singlet excitons (ps---ns). 
This negates Mechanism 1, since Eq.~\eqref{eq:mechanism_1_1} does not apply. 
Nevertheless, it is worth mentioning that ultrafast (picosecond) ISC rates, reported by \citet{alias2021ultrafast}, would lead to more efficient energy transfer to the storage layer.
Recent progress on controlled-ISC can bring a new dimension to our proposed triplet-based energy storage approach.  
Phosphorescence can be activated thermally~\cite{yang2017recent} and optically~\cite{fleischer2017optically}. 
Whenever the activation energy does not exceed the triplet energy, these methods have the potential to further limit losses and open to on-demand work extraction. 
Another important aspect that has been investigated in recent years is the mechanism behind reverse intersystem crossing (RISC) in similar strongly-coupled organic molecular systems \cite{Mukherjee_JACS2023,Eizner_SA2019,Yu_NC2021,Stranius_NC2018,Liu_AP2020a}.
Kinetic studies show that as the LP energy approaches the triplet energy, the decay rate of the triplet state is enhanced, which is attributed to an increased ISC rate \cite{Stranius_NC2018,Mukherjee_JACS2023}.
Here, instead, we suggest that the shorter triplet lifetime is the result of the triplet-polariton hybridisation that occurs when triplets have a non-vanishing singlet character (potentially due to molecular aggregation in the storage layer).
Additionally, by inverting the energy of the LP and triplet state (i.e. a negative detuning, by our nomenclature) in thermally-activated delayed fluorescence materials, it is shown that the RISC rate is unaffected~\cite{Eizner_SA2019} when compared to outside the cavity, however the RISC thermal activation energy is reduced in so-called \textit{barrier-free} RISC~\cite{Yu_NC2021}. 
Interestingly, metastable triplet states with 1-second lifetimes, realised by \citet{an2015stabilizing}, could lead to a self-discharge time 9-order of magnitude longer than that reported in Ref.~\citet{Quach_SA2022a}.

A challenge for future work is to experimentally characterise the performance of the ISC-mediated ``mechanism 1" (Fig.~\ref{fig:mechanisms}). 
Our numerical modelling (detailed in Sec.~\ref{sm:oqs_model_dynamics} of the SM), suggests that this regime has a far better energy storage performance, vastly outperforming it in terms of stored energy density and longer self-discharge time, while suffering only marginally in terms of charging power. 
In this regard, an important question is to determine if this mechanism is compatible with superabsorption, or if it would instead become a bottleneck for the effective charging power of the device. 
Another interesting aspect is that cavity-exciton interactions in this regime could further enhance the self-discharge times by preventing triplet relaxation pathways, as suggested the modelling shown in Sec.~\ref{sm:oqs_model_dynamics} of the SM.

\section{Conclusions}
\label{s:conclusion}

In this letter we proposed an approach to improve the energy storage lifetime of Dicke QBs by several orders of magnitude, 
by introducing an energy storage layer based on  molecular triplet states. 
We have shown that there are at least two working regimes for these type of QBs, presenting  experimental data in support of an optically-driven mechanism discussed in Sec.~\ref{ss:mechanism_2}. 
To identify these regimes and analyse the emission data, we used an open quantum system model based on an effective Jaynes-Cummings Hamiltonian. 
A key outlook is to extend this model to genuine many-body systems, to study and optimise the performance of these devices. 
For the latter, we propose to consider three key energy storage figures of merit connected to the properties of the devices that we have also considered here: 
(1) The rate at which triplets are populated, which is associated with the \textit{charging power}, 
(2) the steady-state triplet population, which represents the \textit{stored energy density} in the layer, and 
(3) the triplet relaxation timescale, which is linked to the \textit{self-discharge time} of the device. 

Finally, we would like to point at the other outstanding challenges for Dicke quantum batteries, beyond the specific organic mircocavity platform. Firstly, energy density is a great cross-platform challenge for energy storage in quantum systems. 
The eV band-gap offered by excitonic systems can in principle deliver energy densities compatible with optoelectronic devices~\cite{Campaioli_2023a}. However, in practice, random inter-site interactions (here given by molecular aggregation at high dye concentration) tend to limit energy density and charging power of Dicke QBs.
Another interesting research direction is to experimentally study Dicke QBs in the lossless, coherent (or \textit{unitary}) regime typical of ``genuine'' quantum technologies based on entanglement. 
In this context, triplet states can be populated coherently~\cite{collins2023quintet}, for example by means of microwave driving, as done in electron spin resonance experiments~\cite{maune2012coherent}. 
This challenging avenue could be explored on ``cold'' platforms such neutral atoms~\cite{adams2019rydberg}, trapped ions~\cite{bruzewicz2019trapped}, and superconducting systems~\cite{blais2021circuit}.
Indeed, there are numerous approaches to investigate, any of which could potentially result in the demonstration of a first fully-operational QB.

\vspace{5pt}

\begin{acknowledgments}
F.C. thanks G.M. Andolina for insightful discussion. F.C. acknowledges that results incorporated in this standard have received funding from the European Union Horizon Europe research and innovation programme under the Marie Sklodowska-Curie Action for the project SpinSC. P.R. acknowledges support through an Australian Research Council DECRA Fellowship (grant no. DE200100279) and an RMIT University Vice-Chancellor’s Senior Research Fellowship. This work was performed in part at the RMIT Micro Nano Research Facility (MNRF) in the Victorian Node of the Australian National Fabrication Facility (ANFF).
\end{acknowledgments}

\bibliography{main}

\newpage
\clearpage
\newpage

\appendix

\onecolumngrid

\begin{center}
\textbf{\large Supplemental Material for \\ ``\thetitle''}
\end{center}
\setcounter{equation}{0}
\setcounter{figure}{0}
\setcounter{table}{0}
\setcounter{page}{1}
\renewcommand{\theequation}{S\arabic{equation}}
\renewcommand{\thefigure}{S\arabic{figure}}
\renewcommand{\bibnumfmt}[1]{[S#1]}
\renewcommand{\citenumfont}[1]{#1}

\section{Open quantum system model of the microcavity dynamics}
\label{sm:oqs_model_dynamics}
\noindent

Let $\omega_D$ ($\omega_A$) be the average energy of the donor (acceptor) singlet excited state, and $\omega_T$ the average energy of the triplet states of the acceptor molecules. 
To describe the system we use an effective model where the donor (acceptor) ensemble is replaced by a single system with cumulative coupling $J_D = \sum_{j=1}^{N_D} g_D^{(j)}$~\cite{Tibben_CR2023},
\begin{equation}
	\label{eq:Rabi_model}
	H = \omega_C(\theta)a^\dagger a + \sum_{\nu = D,A} \omega_\nu \sigma_+^{(\nu)}\sigma_-^{(\nu)} + J_\nu \sigma_x^{(\nu)} (a+a^\dagger),
\end{equation}
where $a$, $a^\dagger$ are the annihilation and creation operators of the cavity and $\sigma_\pm^{(\nu)} = \sigma^{(\nu)}_x\pm\sigma^{(\nu)}_-$ are the singlet-exciton creation and annihilation operator at chromophore $\nu = D,A$. Note that the acceptor $A$ is a three-level system since it can also host a triplet exciton.
If the exciton-photon couplings are small enough compared to the other energy scales, i.e., $J_D,J_A \ll \omega_C(\theta),\omega_D,\omega_A$, we can also apply the rotating wave approximation to eliminate the counter rotating terms arising from the exciton-cavity couplings, obtaining
\begin{equation}
	\label{eq:Jaynes-Cummings_model}
	H = \omega_C(\theta)a^\dagger a + \sum_{\nu = D,A} \omega_\nu \sigma_+^{(\nu)}\sigma_-^{(\nu)} + J_\nu \sigma_+^{(\nu)}a + \sigma_-^{(\nu)}a^\dagger.
\end{equation}
The latter is the effective Jaynes-Cummings model associated to the underlying many-body Dicke model in which each donor and acceptor molecule in the ensembles is treated as an individual system.

Here, we focus on the dynamics of the single-excitation manifold for the chromophores (i.e., each molecule can host at most one exciton), cutting off the cavity space to $n_\mathrm{max}$ photons. The system basis reads
\begin{equation}
	\label{eq:basis}
	\mathcal{B} = \big\{|n, \alpha_D, \alpha_A\rangle, \;\; n = \{0,1,\dots,n_\mathrm{max}\}, \alpha_D = \{S_0,S_1\}, \alpha_A = \{S_0,T_1,S_1\}\Big\},
\end{equation}
according to the following short-hand notation for the ground state reads
\begin{equation}
	\label{eq:ground_diabatic}
	|G\rangle := |0,S_0,S_0\rangle:=|0\rangle_C\otimes|S_0\rangle_D\otimes|S_0\rangle_A.
\end{equation}
To study the formation and relaxation of molecular triplets we use a phenomenological Lindblad quantum master equation based on the effective Jaynes-Cummings model of Eq.~\eqref{eq:Jaynes-Cummings_model},
\begin{equation}
	\label{eq:Lindblad}
	\dot{\rho} = -\frac{i}{\hbar}[H,\rho] + \sum_{k} \bigg(\widetilde{L}_k\rho \widetilde{L}_k^\dagger -\frac{1}{2}\{\widetilde{L}_k^\dagger \widetilde{L}_k,\rho\} \bigg), 
\end{equation}
where $\rho$ is the density operator of the system $\rho(t)$ associated with the cavity, donor, and acceptor, $H$ is the model Hamiltonian, and $\widetilde{L}_k = \sqrt{\gamma_k}L_k$ are the Lindblad operators associated to all the incoherent transitions that occur at rate $\gamma_k$. We assume that the system is initially found in its ground state $\rho_0 \approx |G\rangle\!\langle G|$. Emission and absorption experimental results indicate that the laser pump (continuous wave green laser pump, off-resonance from the cavity) excites mostly the donor-cavity polaritons. Over time-scales much longer than the characteristic cavity-donor coupling $\hbar/J_D$ pumping can be modelled as an incoherent process, due to the lossy, room-temperature settings of the experiment. Accordingly, we model absorption by means of \textit{incoherent pumping} into the cavity $L_p$,
\begin{equation}
	\label{eq:incoherent_pumping}
	L_p = a^\dagger\otimes\mathbb{1}_D\otimes\mathbb{1}_A \;\; \text{at rate} \; \gamma_p, \; \;\;\;\text{(\textit{incoherent pumping})}.
\end{equation}
We also model the incoherent radiative emission losses of cavity and chromophores, as well as the non-radiative triplet exciton recombination, i.e., internal conversion (IC),

\begin{align}
	\label{eq:incoherent_cavity_loss}
	& L_C = a\otimes\mathbb{1}_D\otimes\mathbb{1}_A, \;\; \text{at rate} \; \gamma_C, \; \;\;\;\text{(\textit{cavity loss})}, \\
	\label{eq:incoherent_donor_loss}
	& L_D = \mathbb{1}_C \otimes\sigma_-^{(D)}\otimes\mathbb{1}_A, \;\; \text{at rate} \; \gamma_D \;\;\; \text{(\textit{donor loss})}, \\
	\label{eq:incoherent_acceptor_loss}
	& L_A = \mathbb{1}_C\otimes\mathbb{1}_D\otimes\sigma_-^{(A)}, \;\; \text{at rate} \; \gamma_A \;\;\; \text{(\textit{acceptor loss})},
	\\
	\label{eq:incoherent_internal_conversion}
	& L_{\mathrm{IC}} = \mathbb{1}_C\otimes\mathbb{1}_D\otimes |S_0\rangle_A\!\langle T_1|_A, \;\; \text{at rate} \; \gamma_{\mathrm{IC}} \;\;\; \text{(\textit{triplet internal conversion})}.
\end{align}

Finally, we also model incoherent singlet-to-triplet transition within the acceptor, i.e., ISC via
\begin{equation}
	\label{eq:ISC}
	L_{\mathrm{ISC}} = \mathbb{1}_C\otimes\mathbb{1}_D\otimes |T_1\rangle_A\!\langle S_1|_A, \;\; \text{at rate} \; \gamma_\mathrm{ISC}, \; \;\;\;\text{(\textit{intersystem crossing})}.
\end{equation} 
A summary of all model operators and their meaning can be found in Tab.~\ref{tab:model}.

\begin{table}[ht]
		\caption{Summary of the model used for Eq.~\eqref{eq:Lindblad}. Each timescale is indicative of the considered processes. Specific values for the parameters are reported in the figures.}
	\begin{tcolorbox}[tabulars*={\renewcommand\arraystretch{1.2}}%
		{l|l|l},adjusted title=flush left,halign title=flush left,
		boxrule=0.5pt,title = {\textbf{Model summary}}]
		\textit{Process} & \textit{Operator} &  \textit{Timescale} \\
		\hline\hline
		Dipole-dipole interactions \textit{Jaynes-Cummings model} of Eq.~\eqref{eq:Jaynes-Cummings_model} & $H$ & $\hbar J_A^{-1}$, $\hbar J_D^{-1} \approx $ fs --- ps  \\
		\hline
		Incoherent pumping into the cavity & $\sqrt{\gamma_p} L_p$ & $\gamma_p^{-1} \approx$ ns  \\
		\hline
		Cavity radiative recombination loss & $\sqrt{\gamma_C} L_C$ & $\gamma_C^{-1} \approx$ ps  \\
		\hline
		Donor singlet radiative recombination loss & $\sqrt{\gamma_D} L_D$ & $\gamma_D^{-1}\approx$ ns  \\
		\hline
		Acceptor singlet radiative recombination loss & $\sqrt{\gamma_A} L_A$ & $\gamma_A^{-1} \approx $ ns  \\
		\hline
		Acceptor triplet recombination loss (\textit{internal conversion}) & $\sqrt{\gamma_\mathrm{IC}} L_\mathrm{IC}$ & $\gamma_\mathrm{IC}^{-1} \approx$ $\mu$s   \\
		\hline
		Acceptor singlet-triplet transition (\textit{intersystem crossing}) & $\sqrt{\gamma_\mathrm{ISC}} L_\mathrm{ISC}$ & $\gamma_\mathrm{ISC}^{-1} \approx $ ns  
	\end{tcolorbox}
	\label{tab:model}
\end{table}

\subsection{Triplet formation}
\label{ss:triplet_formation}

\begin{figure}
	\centering
	\includegraphics[width=\textwidth]{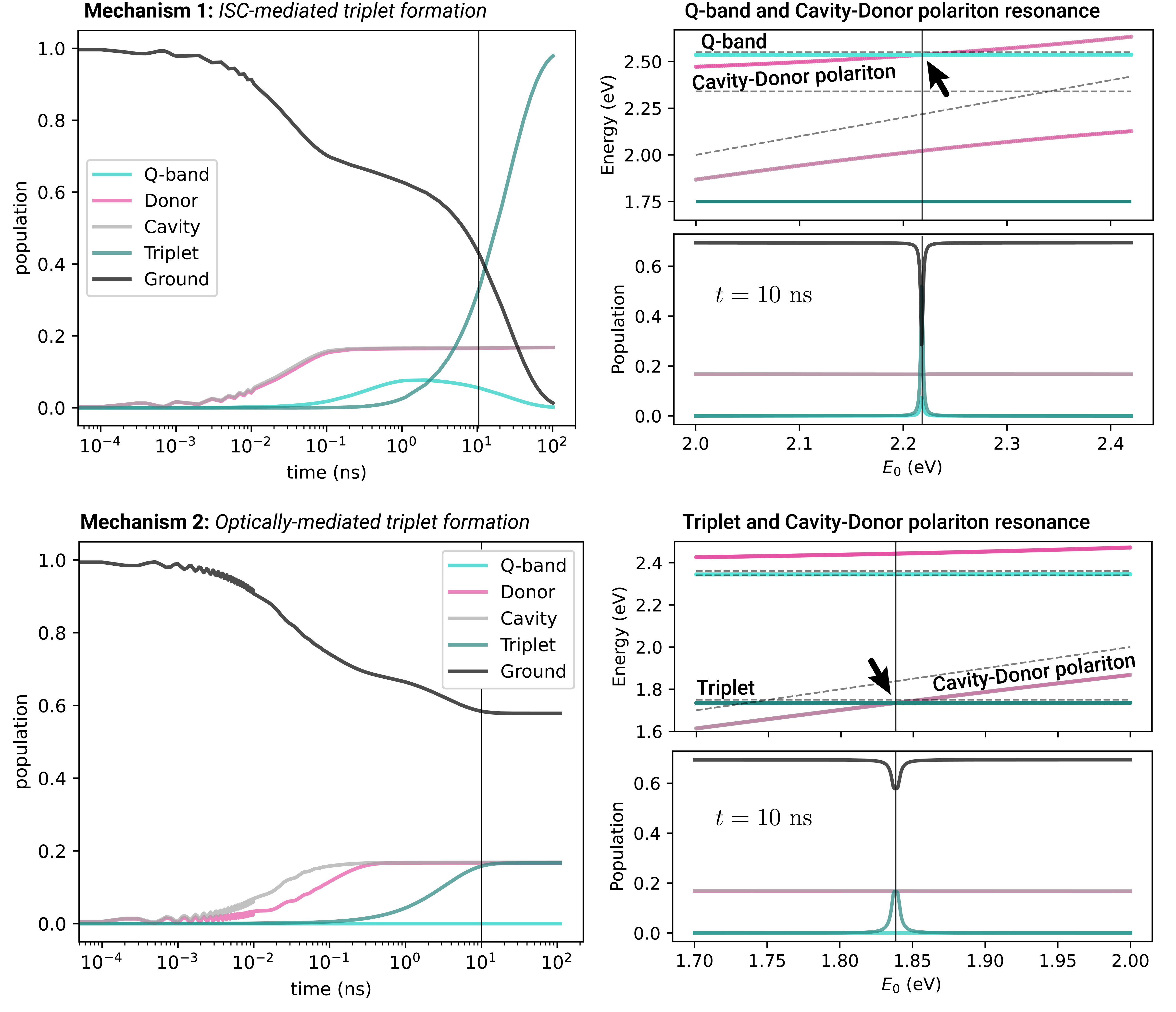}
	\caption{\textbf{Mechanism 1 (\textit{Top}).}---(\textit{Left}) Transient dynamics showing triplet formation in the ISC-mediated mechanism. (\textit{Right}) by sweeping the cavity bare energy $\omega_0$ we cross the resonance between a bright polariton, a cavity-donor hybrid, and the dark  polariton, mostly given by the $S_1$ acceptor state (Q-band). Cavity and donor rapidly reach steady state, while the triplet populates by depleting the ground state via intermediate population of the $S_1$ acceptor state.   \textbf{Mechanism 2 (\textit{Bottom}).}---(\textit{Left}) Transient dynamics showing triplet formation in the optically-mediated mechanism. (\textit{Right}) By sweeping the cavity bare energy $\omega_0$ a resonance is crossed, between the lower bright polariton, a cavity-donor hybrid, and the triplet state. The sharpness of the resonance feature depends on the coupling strength between the resonance states ($J_A$ in mechanism 1 and $J_T$ in mechanism 2), i.e., the weaker the coupling the sharper the resonance. The triplet formation rate depends on the square of this coupling, therefore weaker couplings lead to slower triplet formation. The resonance sweeps are calculated at $t = 10$ ns. The parameters used for these calculations are given in Tabs.~\ref{tab:device_parameters} and~\ref{tab:operation_parameters}.
	}
	\label{fig:results_transfer}
\end{figure}

Triplet formation results are shown in Fig.~\ref{fig:results_transfer}. Here, we compare and discuss the two mechanisms introduced in Sec.~\ref{s:storing_energy_in_triplets} by simulating different devices in the same regime of operation. The device parameters used for each simulation are summarised in Tab.~\ref{tab:device_parameters}, while the operation parameters are given in Tab.~\ref{tab:operation_parameters}. In both cases, triplet formation is studied by applying $\gamma_p = 10$~Ghz pump rate to the cavity, while being mediated by weak 0.1~meV coupling between states that meet a resonance condition.

\begin{table}[ht]
	\caption{Device parameters used for Eq.~\eqref{eq:Jaynes-Cummings_model} to model the properties of the devices discussed in Figs.~\ref{fig:results_transfer} and~\ref{fig:mechanisms}. Here, $J_T$ represents the optical dipole-dipole coupling between the acceptor triplet excitons $T_1$ and the cavity. While we chose completely dark triplet ($J_T = 0$) to illustrate mechanism 1, the same result holds for $J_T \approx 0.1$~meV if the cavity-donor polaritons are sufficiently off-resonance from the triplet state.}
	\begin{tcolorbox}[tabulars*={\renewcommand\arraystretch{1.2}}%
		{l|l|l|l|l|l|l|l},adjusted title=flush left,halign title=flush left,
		boxrule=0.5pt,title = {\textbf{Device parameters}}]
		& $\omega_0$ & $\omega_D$ & $\omega_A$ & $\omega_T$ & $J_D$ & $J_A$ & $J_T$ \\
		\hline\hline
		\textbf{Mechanism 1 (ISC)} & 2.217 eV & 2.34 eV & 2.55 eV & 1.75 eV & 0.25 eV & 0.1 meV & 0.0 meV  \\
		\hline
		\textbf{Mechanism 2 (Optical)} & 1.840 eV & 2.34 eV & 2.36 eV & 1.75 eV & 0.25 eV & 0.1 meV & 0.1 meV  \\
		\hline
	\end{tcolorbox}
	\label{tab:device_parameters_SM}
\end{table}

\begin{table}[ht]
	\caption{Parameters used for Eq.~\ref{eq:Lindblad} to model the operation of both devices in Figs.~\ref{fig:results_transfer} and~\ref{fig:mechanisms}. }
	\begin{tcolorbox}[tabulars*={\renewcommand\arraystretch{1.2}}%
		{l|l|l|l|l|l|l},adjusted title=flush left,halign title=flush left,
		boxrule=0.5pt,title = {\textbf{Operation parameters}}]
		& $\gamma_\mathrm{ISC}$ & $\gamma_D$ & $\gamma_A$ & $\gamma_C$ & $\gamma_\mathrm{IC}$ & $\gamma_p$ \\
		\hline\hline
		\textbf{Pumping and losses for mechanisms 1 and 2} & 0.48 Ghz & 1 Ghz & 1 Ghz & 50 Ghz & 0.1 Mhz & 10 Ghz \\
		\hline
	\end{tcolorbox}
	\label{tab:operation_parameters}
\end{table}

In mechanism 1, shown in Fig.~\ref{fig:results_transfer} (\textit{top}), triplet formation proceeds via the intermediate population of the polariton that has predominately acceptor character. A continuous pumping of the system results in a complete depletion of the ground state, followed by the complete charging of the triplet state. While the triplet formation rate is comparable in both cases, mechanism 1 leads to the optimal steady-state energy density in the system
\begin{equation}
	\label{eq:steady_state_energy_density_1}
	\mathcal{E}(\infty) = \lim_{t\to\infty} \omega_{T}\sum_{i=1}^{N_D} \frac{p^{(i)}_T(t)}{N_D} = \omega_{T} p_T(\infty),
\end{equation}
where $p_T^{(i)}(t)$ is the population of each individual triplet molecular state at time $t$, $N_D$ is the number of donor molecules, and $p_T(\infty) = \mathrm{Tr}[\rho(\infty) \mathbb{1}_C \otimes \mathbb{1}_D\otimes |T_1\rangle_A\langle T_1|_A]$ is the steady-state average triplet population.

In contrast, mechanism 2, shown in Fig.~\ref{fig:results_transfer} (\textit{bottom}), leads to triplet formation via the direct optical dipole-dipole coupling between the bright triplet and the cavity. Triplet formation is the fastest when the lower cavity-donor polariton is at resonance with the triplet energy. In the optimal resonance conditions, this mechanism leads to a triplet formation rate that is about 1 order of magnitude faster than that of mechanism 1. However, a key consequence of this direct coupling mechanism is that the triplets are hybrid polariton states, and thus inherit a shorter lifetime due to the cavity, donor and singlet acceptor fast radiative recombination losses. This leads to a sub-optimal steady-state stored energy density $\mathcal{E}(\infty)$, that is significantly lower than for mechanism 1. 

\subsection{Triplet relaxation}

Here, we simulate triplet relaxation by switching off the pumping after a nominal ``charging time'' of 100~ns, i.e., sufficiently long for both devices to have reached steady-state conditions. The results are shown in Fig.~\ref{fig:mechanisms}. As anticipated in Sec.~\ref{ss:mechanism_1}, mechanism 1 leads to long-lived triplets, while mechanism 2 is affected by rapid relaxation, which is comparable to the Thz--Ghz radiative emission rate of cavity and singlet excitons. Interestingly, mechanism 1 performs even better than anticipated, since the resulting triplets cannot fully decay via direct non-radiative recombination (internal conversion, IC) of the bare triplet. This is a special condition that only occurs due to the non-vanishing $J_D$ and $J_A$ couplings, which shift the system's ground state enough to partially forbid internal conversion (IC). As a result, in the absence of other recombination pathways, the triplets formed this way are trapped in the device and live indefinitely. As shown in Fig.~\ref{fig:mechanisms} (\textit{top}), the triplets lose only a fraction of their population via IC, remaining significantly populated.

To show the effect of cavity-exciton coupling on the self-discharge process, we also study triplet relaxation by switching off pumping and ``opening'' the cavity, i.e., switching off the $J_D$ and $J_A$ couplings. The results, illustrated in Fig.~\ref{fig:mechanisms} (dashed lines), show that relaxation proceeds following standard IC when triplets are bare. Another important aspect of mechanism 2 is that it can in principle directly benefit from superabsorption, being directly mediated by optical couplings. In this case, a way to prevent fast relaxation is to bring cavity and triplets sufficiently off-resonance to restore the bare triplet lifetime ($1/\gamma_\mathrm{IC}$).

\section{Device fabrication details}
\label{sm:fabrication_details}

\begin{figure}
    \centering
    \includegraphics[width=0.45\linewidth]{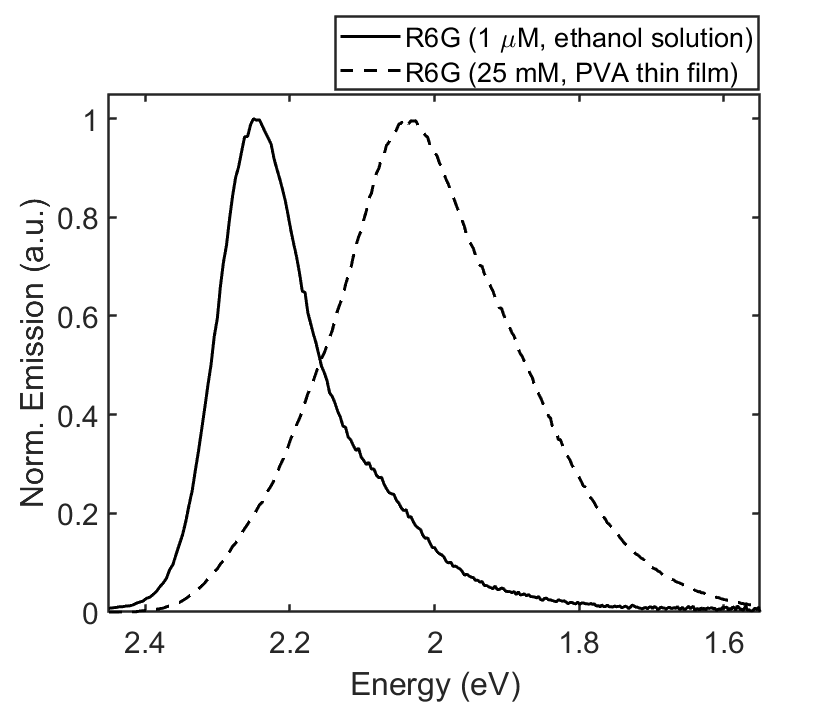}
    \caption{Red-shifted R6G emission due to molecular aggregation in polymer thin films.}
    \label{fig:R6G-emission}
\end{figure}

Firstly, onto an evaporated 100 nm silver film on silicon, a solution of 5 mM PdTPP in 2wt.\% polymethyl methacrylate (PMMA) in chlorobenzene was deposited as the storage layer.
This was followed by two subsequent polymer film depositions of 1wt.\% polyvinyl alcohol (PVA) in a 1:1 ethanol/water solvent mix, and 1wt.\% PMMA in chlorobenzene to create the polymer spacer layer.
A solution of 25 mM R6G in 2wt.\% PVA in ethanol/water as then deposited as the charging layer.
The microcavity was completed by deposition of a 25 nm thin film of evaporated silver. R6G, at high concentrations in thin--films, shows a broadened and red--shifted emission due to molecular aggregation~\cite{Levshin_JAS1977,Bojarski_CPL1997} (Fig. \ref{fig:R6G-emission}).

\section{Additional measurements}
\label{sm:measurments}

\subsection{Reflectance--absorption and emission dispersion}
Angle--resolved reflectometry was used to ascertain the absorption--dispersion characteristics of the microcavity devices.  
Acquired with Agilent Cary 7000 UV-Visible spectrophotometer with Universal Measurement Accessory (UMA) and xenon lamp source.
Angle--resolved emission was acquired by back focal plane imaging (NA = 0.45) through a Bertrand lens and spectrometer, with 514 nm CW laser source at $\sim1.6 \mu$W.
\begin{figure*}[th!]
	\centering
	\includegraphics[width=0.9\textwidth]{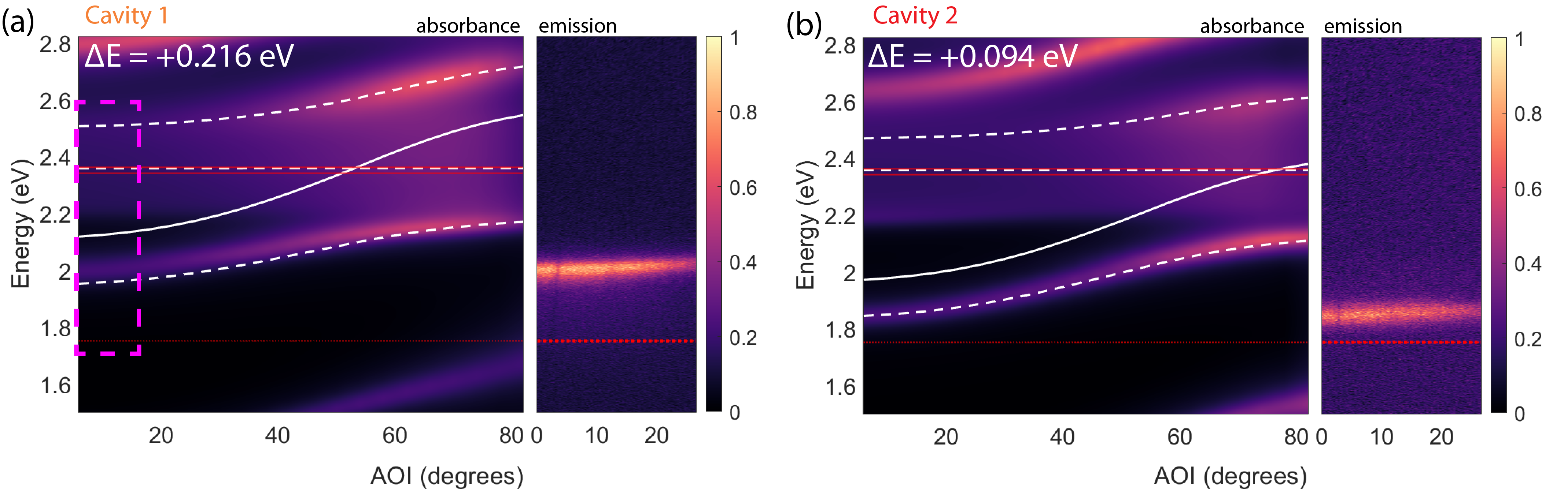}
    \includegraphics[width=0.9\textwidth]{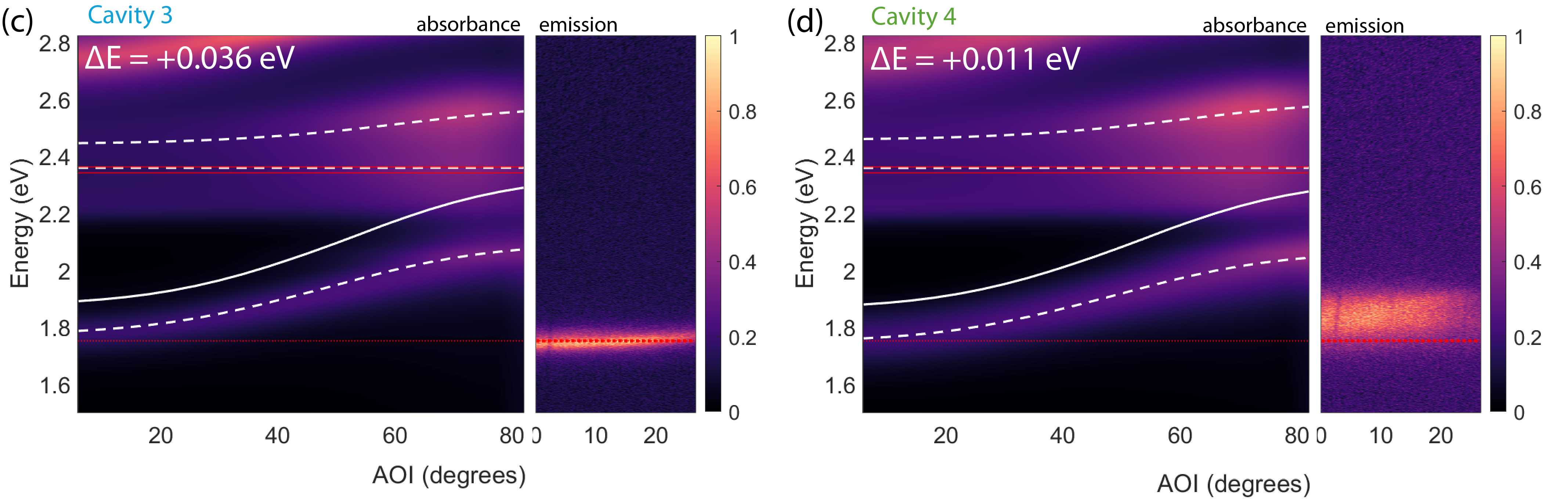}
    \includegraphics[width=0.9\textwidth]{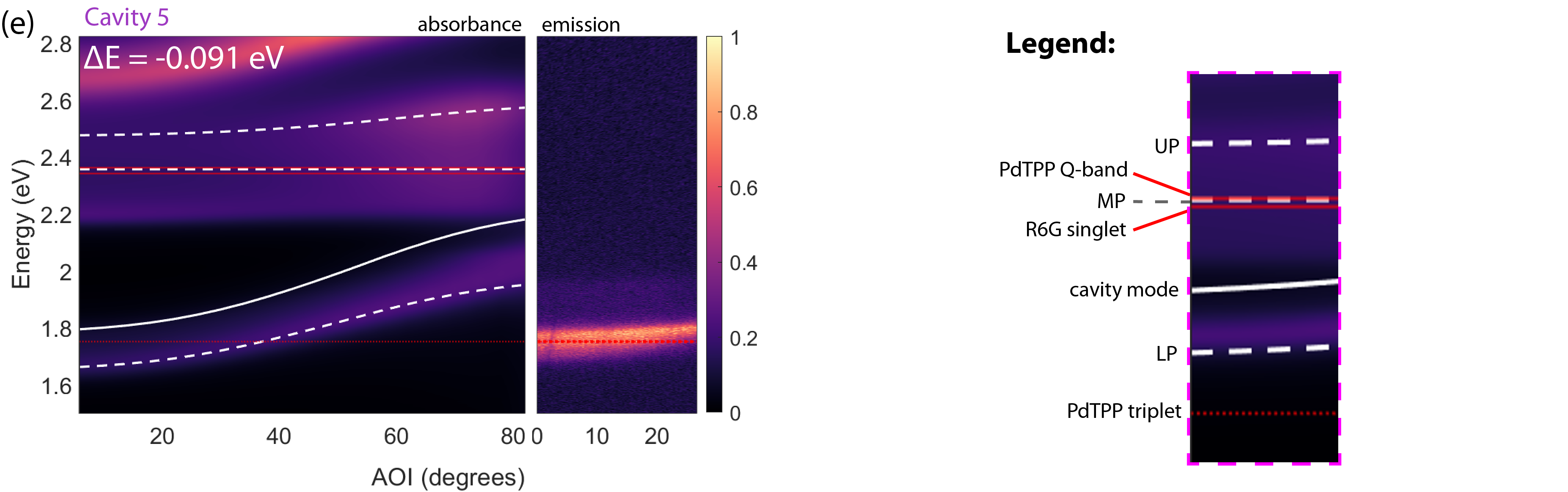}
	\caption{\small{Measured absorption and emission of microcavity devices.
 These are shown for multiple lower polariton--triplet energy detunings $\Delta E = E_{LP} - E_\mathrm{T_1} $, showing detuning--dependent emission.
 All plots are normalised to their maximum intensities.
		(a) Cavity 1, $\Delta E =$ +0.216 eV. Large positive detuning results in emission primarily from the LP. 
		(b) Cavity 2, $\Delta E =$ +0.094 eV. Moderate positive detuning results in emission primarily from the LP. 
		(c) Cavity 3, $\Delta E =$ +0.036 eV. Near--isoenergetic detuning results in emission around the triplet energy. 
		(d) Cavity 4, $\Delta E =$ +0.011 eV. Near--isoenergetic detuning, in this case, results in broadened emission around the triplet energy. 
		(e) Cavity 5, $\Delta E =$ -0.091 eV. Negative detuning results in emission primarily from the LP.
        }}
	\label{fig:absorption-emission}
\end{figure*}
\subsection{Phosphorescence intensity} 
\begin{figure*}[th!]
    \centering
    \includegraphics[width=0.9\textwidth]{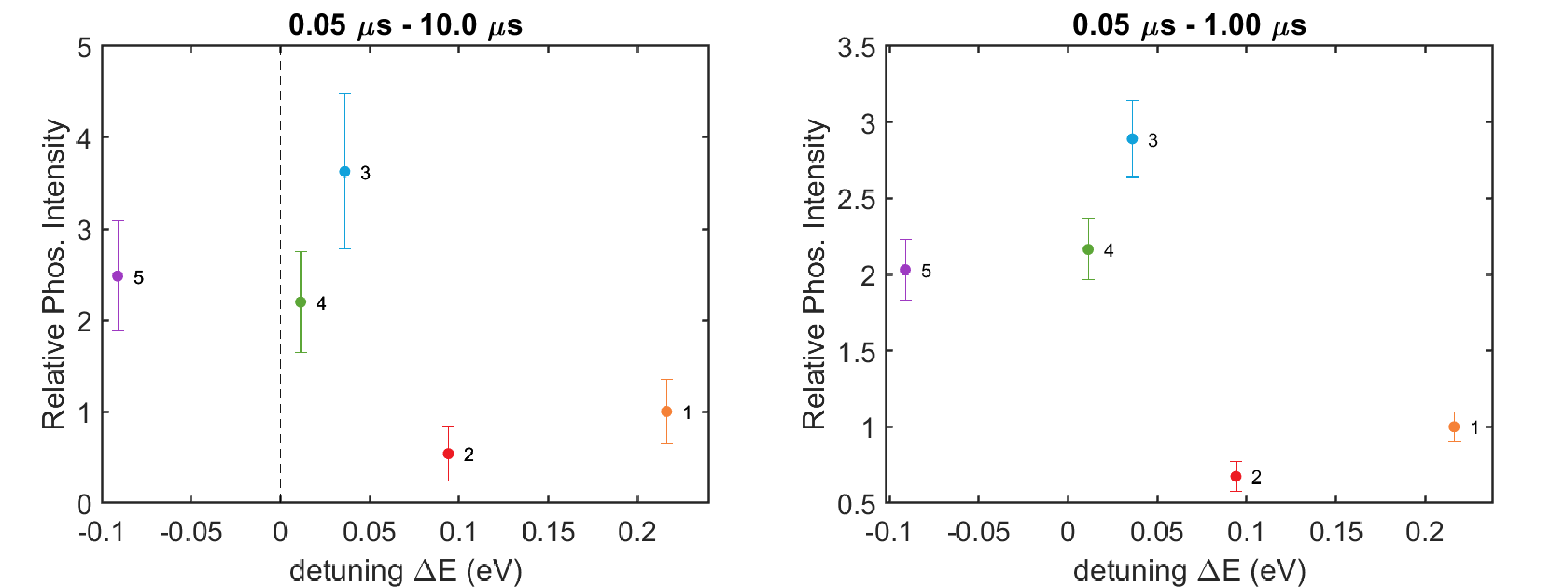}
    \caption{Relative intensity of phosphorescence with respect to Cavity 1, measured at (\textit{Left}) 0.05 $\mu$s to 10.00 $\mu$s and (\textit{Right}) 0.05 to 1.00 $\mu$s emission delay. The vertical dashed line is representative of $\Delta E = 0$, and the horizontal dashed line is representative of the normalisation reference point (Cavity 1 phosphorescence intensity). }
    \label{fig:phosphorescence}
\end{figure*}

Time-resolved photoluminescence spectroscopy was used to ascertain the phosphoresence lifetime and intensity for the devices, with a resolution of  12.8 ns, maximum timescale of 10.0 $\mu$s and incident 520 nm laser power of $\sim$50 nW. 
Relative intensity of phosphorescence across the devices was measured at time 0.05 - 10.0 $\mu$s and 0.05 - 1.00 $\mu$s by taking the area under the curve with the trapezoidal numerical integration method outlined in Eq. \ref{eq:trapz}, and normalised to the peak intensity of Cavity 1 (see Fig. \ref{fig:phosphorescence}). 
\begin{equation}
    \label{eq:trapz}
        \int_{a}^{b} f(x) dx \approx \frac{b-a}{2N} \sum_{n=1}^{N}(f(x)+f(x_{n+1}))
\end{equation}
These data show a maximum of phosphorescence intensity at Cavity 3, as $\Delta E \rightarrow 0$.
We expect the phosphorescence intensity of Cavity 4 over this timescale is reduced due to its shortened ``lifetime'' ($\gamma_\textrm{ph}^{-1}$), and when measured at shorter timescales, Cavity 4 phosphorescence relative intensity increases.
Other factors, such as minor differences in device fabrication (e.g. dissolved oxygen levels, cavity quality factor) may also contribute to uncertainty when comparing phosphorescence intensity across the devices.

\end{document}